\documentstyle[epsfig,amsfonts]{JHEP}
\title{Topological defects in lattice gauge theories}
\author{A.C. Davis \\ DAMTP, CMS, Wilberforce Rd, Cambridge, CB3 0WA, U.K.
\\ \email{a.c.davis@damtp.cam.ac.uk}
}
\author{T.W.B. Kibble 
\\ Blackett Laboratory, Imperial College, London SW7 2BW, U.K.
\\ \email{t.kibble@ic.ac.uk}}
\author{A. Rajantie
\\ Centre for Theoretical Physics,
University of Sussex, \\Falmer, Brighton BN1 9QJ, U.K.
\\ \email{a.k.rajantie@damtp.cam.ac.uk}}
\author{H. Shanahan
\\Center for Computational Physics, University of Tsukuba, \\Tsukuba, 
Ibaraki 305-8577, Japan\\
\email{shanahan@rccp.tsukuba.ac.jp}}
\abstract{
We present a non-perturbative formalism for measuring defect free 
energies (monopole mass or vortex tension) in three-dimensional
SU(2)+adjoint Higgs models. 
Starting from twisted, translation invariant boundary conditions, we
perform a change of variables that allows us to express the defect
free energies in terms of 't~Hooft loops.
We propose that the defect free energies can be used to distinguish between
phases in this model,
and also more generally in other gauge field theories where no
local order parameters exist.
In the case of monopoles, our construction can also be used
in four-dimensional pure-gauge SU(2) theory, where it gives
the monopole mass 
in the maximally Abelian gauge without 
the need of actually fixing the gauge in the simulation.
Moreover, the expression is manifestly independent of the choice
of the Abelian projection as long as it is
compatible
with the classical
't~Hooft-Polyakov solution.
}
\preprint{
\begin{tabular}{l}
DAMTP-2000-108\cr
Imperial/TP/99-0/43\cr
SUSX-TH-00-013\cr
hep-lat/0009037
\end{tabular}
}
\begin{document}
\def\ph{\hat\Phi}
\def\ch{\hat\chi}
\def\U{{\mathbb I}}
\def\tch{\tilde\chi}
\def\tU{\tilde U}
\def\tV{\tilde V}
\def\tL{\tilde\Lambda}
\def\x{\vec{x}}
\def\xpmu{\x\!+\!\hat\mu}
\def\xpi{\x\!+\!\hat\imath}
\def\xpj{\x\!+\!\hat\jmath}
\def\xpij{\x\!+\!\hat\imath\!+\!\hat\jmath}
\def\tr{{\rm Tr}\,}
\def\nn{\nonumber}
\def\kh{\hat{k}}
\def\jh{\hat\jmath}
\def\ih{\hat\imath}
\def\xh{\hat{x}}
\def\yh{\hat{y}}
\def\zh{\hat{z}}
\def\O{{\mathcal O}}

\section{Introduction}
In many cases, phase transitions can be described in terms of topological
defects. In some simple models, such as the Ising 
model~\cite{Kramers:1941kn} and super-Yang-Mills
theories~\cite{Seiberg:1994rs}, 
an exact relation is known between this dual formulation and
the original formulation of the theory.
However, even without such an exact mapping, this dual description
can be very useful, in particular since it gives a way of finding
natural order parameters for the phase transition.

If the phase transition is associated with a spontaneous
breakdown of a global symmetry, essentially any local operator that 
transforms
non-trivially under the symmetry group can be used as an order parameter.
However, if the broken 
symmetry is a gauge invariance, this approach does not work,
because only gauge-invariant
operators have non-zero expectation values. Although sometimes, 
for instance in the
electroweak theory~\cite{Fradkin:1979dv,Kajantie:1996mn}, 
the transition predicted by perturbation theory
is in fact only a smooth crossover, often a well-defined transition
exists. In these cases, a dual picture in terms of topological defects
can often be the most natural description.

There are two possible ways of constructing an order parameter based on
topological defects. In theoretical studies, it is convenient to
consider the creation operator $\mu$ of a
defect~\cite{Kadanoff:1971kz,Kleinert:1982dz,%
Frohlich:1986sz,%
Kovner:1991pz,DelDebbio:1995sx,Chernodub:1997ps,Frohlich:1999wq}. 
In the low-temperature
phase, its expectation value $\langle\mu\rangle$ vanishes, 
but if the defects condense in the 
high-temperature phase, it becomes non-zero. Thus it behaves as an
ordinary order parameter in a spontaneous symmetry breakdown, but with
an inverted temperature, and often it can indeed be associated with a
symmetry of the dual theory that is broken in the high-temperature 
phase. Therefore it is often called a disorder 
parameter~\cite{Kadanoff:1971kz}.
However, this approach is only useful if the defects are line-like
and can therefore be interpreted as world-lines of particles of the dual
theory.

More generally, one can use the defect free 
energy~\cite{Onsager:1944jn,Kajantie:1999zn,Kovacs:2000sy,Hart:2000en%
,deForcrand:2000fi}
as an
order parameter. In the special case of line-like defects,
the tension, i.e.~free energy per unit length, is
simply the inverted correlation
length of $\mu$, and thus vanishes at the transition point.
Using the defect free energy as the order parameter also avoids the
problem that in numerical simulations $\langle\mu\rangle$
always vanishes because of the finite lattice size.

In the Abelian Higgs model,
Nielsen-Olesen vortices have been used successfully to study the phase 
transition. In Refs.~\cite{Kleinert:1982dz,Kovner:1991pz}, 
the disorder parameter
and the corresponding U(1) symmetry were discussed, and in 
Ref.~\cite{Kajantie:1999zn}
the tension of a vortex was measured
and shown to be an order parameter.
The purpose of this paper is to extend these studies to
SU(2) theories with one or two adjoint Higgs fields
by constructing explicitly the observables that measure
free energies of vortices and monopoles.

Even though there is no spontaneous gauge symmetry breakdown in
pure-gauge SU(2) theory, it is still believed that the non-perturbative
dynamics of the gauge fields can give rise to similar effects.
Both vortices~\cite{'tHooft:1978hy} and monopoles~\cite{Mandelstam:1976pi} 
have been suggested as possible
explanations for confinement. While the vortex tension can be expressed
simply as the expectation value of a 't~Hooft loop~\cite{'tHooft:1978hy}
 and the procedure 
of measuring
it is well 
understood~\cite{Kovacs:2000sy,Hart:2000en,deForcrand:2000fi},
giving a non-perturbative definition for the monopole mass is
more involved.
It requires defining a composite
adjoint field with an Abelian projection~\cite{'tHooft:1981ht}, 
after which 
monopoles can be defined in the same way as in the Georgi-Glashow 
model~\cite{'tHooft:1974qc,Polyakov:1974ek}. 
Disorder parameters based on monopoles
have been studied in 
Refs.~\cite{DelDebbio:1995sx,Chernodub:1997ps,Cea:2000zr},
but they depend on the choice of the Abelian projection,
refer to solutions of classical field equations 
and
don't have the same elegance as expressing the vortex tension
in terms of a 't~Hooft loop.

In this paper, we present a gauge invariant, non-perturbative formalism for
studying vortices and monopoles in SU(2) theories with adjoint Higgs fields. 
In Sect.~\ref{sect:topodef}, we discuss topological defects and 
defect free energies in general terms.
Sect.~\ref{sect:resiu1} contains the derivation of 
a gauge-invariant expression for the 
residual U(1) gauge field in the broken phase.
In Sect.~\ref{sect:2Higgs}, 
we discuss the two-Higgs model and
review the definition of the vortex tension in terms of a 't~Hooft loop.
This acts as an introduction to the definition of the monopole mass
in the Georgi-Glashow model, which is carried out in 
Sects.~\ref{sect:1Higgs}.
In Sect.~\ref{sect:puregauge},
we show how the same construction can be used
in pure-gauge theories with Abelian projections.

\section{Topological defects}
\label{sect:topodef}
In continuum, topologically stable non-trivial solutions of the field
equations, i.e.~topological defects, exist when the vacuum manifold
of the system has non-trivial topology. 
They are characterized by a winding number, which
is non-zero if a defect is present.
The mass of such a defect
can be defined simply as the energy of the field configuration.
When the symmetry is restored, the distinction between a defect and the
vacuum disappears, and we can say that the mass of the defect vanishes.

However, in quantum systems and in classical systems at finite
temperature, 
the state of the system is not given by a single 
field configuration, but rather by a density operator or an ensemble
of configurations. In both cases, a useful description for the state
of the system is given by the partition function, which has the form
\begin{equation}
\label{equ:partfun}
Z=\int D\phi \exp(-S).
\end{equation}
Here $S=\int d^Dx {\cal L}$ 
is the action and in the quantum case, we have performed
the Wick rotation. In the thermal case, $S=\beta H$, 
but for simplicity we assume that
the temperature is absorbed in 
the parameters of the Lagrangian ${\cal L}$.

The partition function is a path integral over a large number of field
configurations, almost 
none of which are solutions of the equations of motions.
For each configuration, the winding number of the whole system is still
a well-defined quantity. When it
is zero, there can be localized objects that at a suitable length
scale have the characteristics of a topological defect, but the total
number of defects and anti-defects (those with a negative winding number)
must be equal, and therefore they can be thought of as defect-antidefect
pairs created by thermal fluctuations. However, if the total winding
is non-zero, the number of defects and anti-defects is not equal, 
which means that the configuration contains, in addition to
the thermally generated defect-antidefect pairs, true topological
defects. Note that in general it is not possible to determine which
of the defects are the
true ones and which are members of the thermally generated
pairs.

From the point of view of local observables, it
does not make any difference in the thermodynamical limit if we 
use instead of the canonical partition function (\ref{equ:partfun})
a microcanonical one $Z_{n}$, where 
the integration is restricted to configurations with
a given total winding number $n$.
However, the value
of $Z_{n}$ is different for each $n$. 
In fact, it is more useful to consider the
free energy
\begin{equation}
F_{n}=-\ln Z_{n},
\end{equation}
because in the classical limit where the integral is dominated by its saddle
point, the free energy difference
\begin{equation}
\label{equ:fediff}
\Delta F=F_1-F_0
\end{equation}
is exactly the classical mass of the
defect. We will call $\Delta F$ the free energy of a defect. Note that 
it is defined without recourse to the saddle-point approximation.

It is not obvious how to use Eq.~(\ref{equ:fediff}) in lattice field theories,
because the topology of the field configuration space is very different
in a discrete space. For instance, the Abelian Higgs model has two
different lattice formulations, the compact and the non-compact one,
and only the non-compact version contains topological defects.
Thus we will first have to show that in any particular case
we are discussing, the winding number can really be defined.
Even that is not enough, because
in lattice simulations, it is only possible to
measure expectation values
\begin{equation}
\langle{\cal O}\rangle=Z^{-1}\int D\phi{\cal O}\exp(-S),
\end{equation}
and not free energies directly.
Therefore, we will also have to rewrite Eq.~(\ref{equ:fediff})
in terms of expectation values.

\section{Residual U(1) invariance}
\label{sect:resiu1}
Let us start by considering in general an SU(2) gauge theory with
an adjoint scalar field $\Phi$, which may be either fundamental or
composite. 
We will now derive an expression for the Abelian gauge field corresponding to
the residual U(1) gauge invariance that remains if the field $\Phi$
breaks the SU(2) symmetry.
A similar construction has previously been carried out
in continuum~\cite{'tHooft:1974qc} and
on a lattice in 
Refs.~\cite{Kronfeld:1987vd,Chernodub:1998bi}.

In this section, we assume that the system is defined on a three-dimensional
lattice, which may also
be a single time slice of a four-dimensional lattice.
The gauge field is represented by SU(2) matrices
$U_i(\x)$  defined on the links $(\x,\xpi)$
between the lattice sites $\x\in\{0,\ldots,N-1\}^3$,
and the adjoint scalar field $\Phi$ is defined on lattice sites.

The theory is invariant under gauge
transformations 
\begin{eqnarray}
\Phi(\x)&\rightarrow&\Lambda^\dagger(\x) \Phi(\x) \Lambda(\x),\nn\\
U_i(\x)&\rightarrow&\Lambda^\dagger(\x) U_i(\x) \Lambda(\xpi),
\end{eqnarray}
where $\Lambda(\x)$ is an SU(2)-valued function defined on the lattice
sites.
In particular, it is always possible to gauge transform $\Phi$ to the
$z$-direction.

To simplify the discussion, we neglect the configurations
in which $\Phi$ either vanishes or is proportional to $\sigma_3$ at any
lattice site. 
Because they have a zero measure in the partition function,
this does not change the results, and our final
results work well even in these special cases.
This allows us to
define the unit vector
\begin{equation}
\ph=\Phi(\Phi^2)^{-1/2}.
\end{equation}

We start the construction by making a gauge transformation that diagonalizes 
$\Phi$, i.e.~turns $\ph$ into $\sigma^3$ at every point.
That is accomplished by
\begin{equation}
\label{equ:Rdef}
 R(\x)\propto i(\sigma^3+\ph(\x)).
\end{equation}
Here and in the following we use 
the notation $\propto$ to show an equality that is
true up to a real and positive factor. 
Our final expressions will be independent of
these factors.
In Eq.~(\ref{equ:Rdef}), the normalization
is chosen in such a way that $R$ is an SU(2) matrix.

We can now define the transformed gauge field
\begin{equation}
\label{equ:unigauge}
\tU_i(\x)= R^\dagger(\x) U_i(\x) R(\xpi).
\end{equation}
Since the corresponding value of $\Phi$ is proportional to $\sigma^3$,
$R$ is nothing but a gauge transformation into the unitary gauge.
However, this does not fix the gauge completely, because of the
residual U(1) invariance. More precisely, a gauge transformation with
a matrix $\Lambda$ induces a transformation
\begin{equation}
\tU_i(\x)\rightarrow\tL^\dagger(\x)\tU_i(\x)\tL(\xpi), 
\end{equation}
where
\begin{equation}
\tL=R^\dagger\Lambda R_\Lambda,
\end{equation}
and
\begin{equation}
R_\Lambda\propto i(\sigma^3+\Lambda^\dagger\ph\Lambda).
\end{equation}

The new transformation $\tL$ is unitary,
\begin{equation}
\tL^\dagger\tL=
R^\dagger_\Lambda\Lambda^\dagger RR^\dagger\Lambda R_\Lambda=1,
\end{equation}
and its determinant
is one.
It is also diagonal, since it leaves $\sigma^3$ invariant by 
construction,
and therefore it must be of the form 
$\tL=\exp(i\lambda\sigma^3)$. This is the residual Abelian gauge
transformation.

Let us then show that the corresponding gauge field is given by
the phase of the diagonal elements of $\tU_i$.
We define the projection operators
\begin{equation}
P_\pm=\frac{1}{2}(\U\pm\sigma^3),\quad \mbox{i.e.}\quad
P_+=\pmatrix{1&0\cr 0&0},\quad P_-=\pmatrix{0&0\cr 0&1},
\end{equation}
and the projected fields
\begin{equation}
\tV_{i,\pm}(\x)=P_\pm\tU_i(\x) P_\pm.
\end{equation}
The field $\tV_{i,+}$ ($\tV_{i,-}$) corresponds to the
upper left (lower right) component of the total gauge field $\tU_i$.
Furthermore, if we define
\begin{equation}
\Pi_\pm=\frac{1}{2}(\U\pm\ph),
\end{equation} 
we can write
$\tV_{i,\pm}(\x)=R^\dagger(\x)V_{i,\pm}(\x)R(\xpi)$,
where
\begin{equation}
V_{i,\pm}(\x)=\Pi_\pm(\x) U_i(\x)\Pi_\pm(\xpi).
\end{equation}

Under a gauge transformation, the fields $\tV_{i,\pm}$ transform as
\begin{equation}
\tV_{i,\pm}(\x)\rightarrow
\tL^\dagger(\x)\tV_{i,\pm}(\x)\tL(\xpi)
=\exp\left\{\pm i\left(\lambda(\xpi)-\lambda(\x)\right)\right\}\tV_{i,\pm}(\x),
\end{equation}
which means that
\begin{equation}
\label{equ:gaugedef}
\alpha_i\equiv
\arg\tr \tV_{i,+}=-\arg\tr \tV_{i,-}
\end{equation}
indeed behaves like the Abelian gauge field.
Thus we can also define the magnetic flux density as
\begin{eqnarray}
\label{equ:fluxdef}
\alpha_{ij}&\equiv&
\arg\tr \tV_{i,+}(\x)\tV_{j,+}(\xpi)\tV^\dagger_{i,+}(\xpj)
\tV^\dagger_{j,+}(\x)\\
&=&
\arg\tr\Pi_+(\x)U_i(\x)\Pi_+(\xpi)U_j(\xpi)
\Pi_+(\xpij)U^\dagger_i(\xpj)
\Pi_+(\xpj)U^\dagger_j(\x)
.\nonumber
\end{eqnarray}
This final version is gauge invariant and contains no reference to
the fields in the unitary gauge.

Using Eq.~(\ref{equ:fluxdef}), we can define the magnetic charge inside 
a lattice
cube as
\begin{equation}
\label{equ:monodef}
C_M(\x)=\frac{1}{4\pi}\sum_{ijk}\epsilon_{ijk}\left(
\alpha_{jk}(\xpi)-\alpha_{jk}(\x)
\right).
\end{equation}
This number is an integer and can be non-zero.

\section{3D two-Higgs model}
\label{sect:2Higgs}
Let us first consider a model with two adjoint Higgs fields,
$\Phi$ and $\chi$.
The Lagrangian is
\begin{eqnarray}
{\cal L}_{\rm 2H}&=&
\frac{4}{ag^2}\sum_{i<j}\left(1-\frac{1}{2}
{\rm Re}\,\tr 
U_i(x)U_j(\xpi)U_i^\dagger(\xpj)U_j^\dagger(x)
\right)\nonumber\\
&&+\sum_{i}2a \biggl[\tr\Phi^2(\x)-\tr\Phi(\x)U_i(\x)
\Phi(\xpi)U_i^\dagger(\x)\nonumber\\
&&\phantom{+\sum 2a[}+\tr\chi^2(\x)-\tr\chi(\x)U_i(\x)
\chi(\xpi)U_i^\dagger(\x)\biggr]\nonumber\\
&&+m_\Phi^2a^3\tr\Phi^2(\x)+a^3\lambda_\Phi
(\tr\Phi^2)^2\nonumber\\
&&+m_\chi^2a^3\tr\chi^2(\x)+a^3\lambda_\chi
(\tr\chi^2)^2+a^3\eta\tr\Phi^2\chi^2
+ a^3 \eta_2 \left(\tr \Phi\chi\right)^2. 
\label{equ:lagr2h}
\end{eqnarray}

\FIGURE{
\epsfig{file=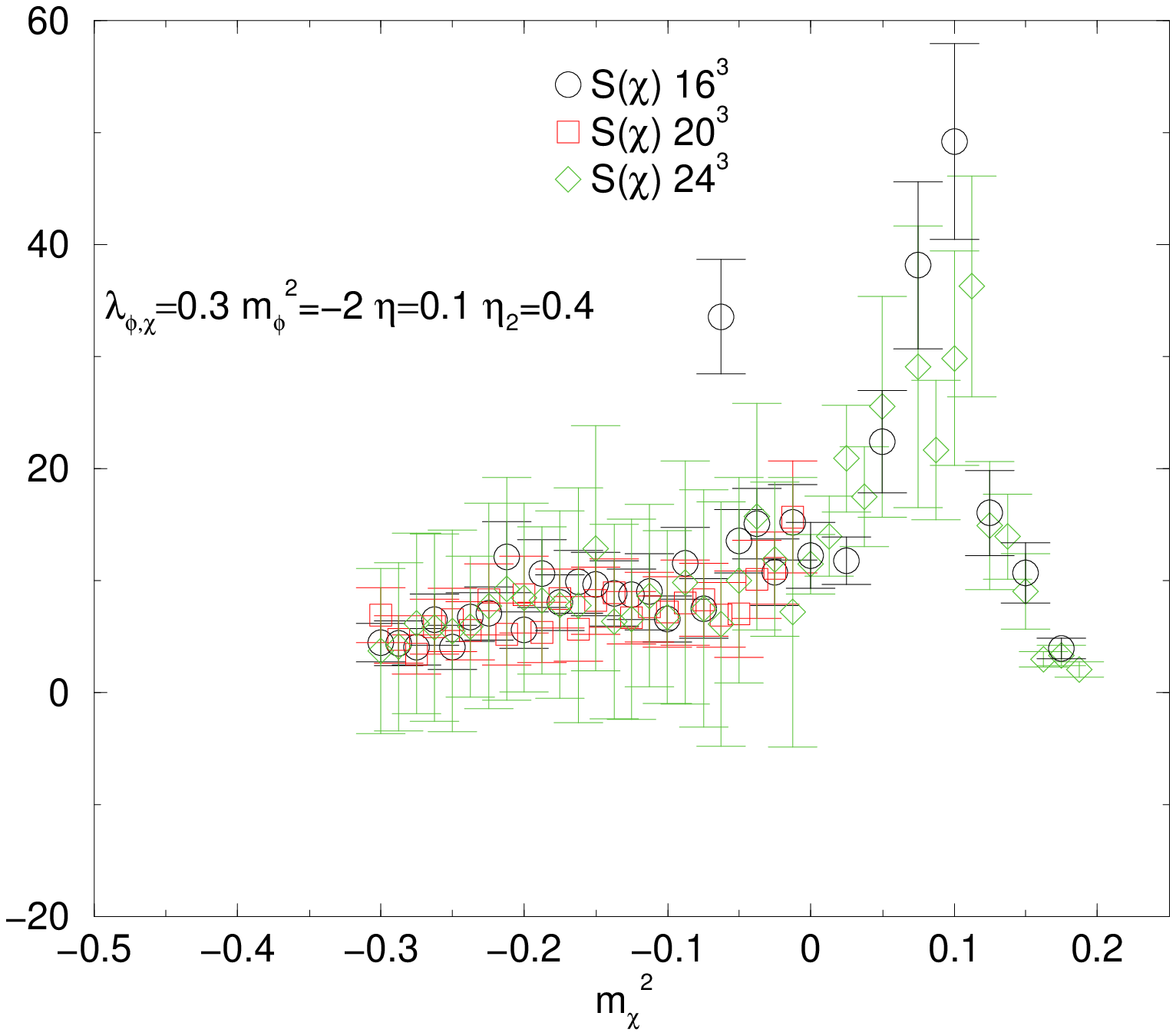,width=10cm}
\caption{The susceptibility of $\tr\chi^2$ near the 
perturbative transition point. }
\label{fig:hugh}
}

If $\Phi$ is non-zero,
the two-Higgs model looks very much
like the Abelian Higgs model. In Fig.~\ref{fig:hugh} 
we show the plot of the
susceptibility of $\tr\chi^2$ at the point
where $\chi$ becomes non-zero in perturbation theory
(see the Appendix for numerical details). 
The maximum susceptibility
does not increase with the volume, and therefore the field $\tr\chi^2$
does not behave non-analytically. The same applies to any local
observable, and therefore we are forced to consider non-local order
parameters, and in particular the defect free energy (\ref{equ:fediff}).
Because in this model, the defects are line-like vortices, 
it is convenient to define the tension as the free-energy per unit length
of a vortex
\begin{equation}
T=\lim_{L\rightarrow\infty}\frac{\Delta F}{L}.
\end{equation}
The construction of this order parameter is well 
known and dates back to late 
seventies~\cite{'tHooft:1978hy,Mack:1979rq}, but we will rederive it here,
because the derivation of the monopole mass in the one-Higgs
model follows similar lines.

\subsection{Winding number}
We start by deriving the lattice expression for the winding number.
We will follow the corresponding derivation in the case of the
Abelian Higgs model~\cite{Kajantie:1998bg} 
(see also Ref.~\cite{Chernodub:1998bi}).
Since we already know the analogue (\ref{equ:gaugedef}) of the
Abelian gauge field, we only have to find
the corresponding Higgs field.
If we define using Eq.~(\ref{equ:Rdef})
\begin{equation}
\tch=R^\dagger\chi R,
\end{equation}
we find that it transforms as
\begin{equation}
\tch\rightarrow \tL^\dagger\tch\tL.
\end{equation}
If we 
now project to the lower left matrix 
element of $\tch$ 
\begin{equation}
h=P_-\tch P_+, 
\end{equation}
we find the transformation law
\begin{equation}
h=P_-R\chi RP_+=R\Pi_-\chi\Pi_+R\rightarrow 
R_\Lambda\Lambda^\dagger\Pi_-\chi\Pi_+\Lambda R_\Lambda
=\tL^\dagger h\tL=\exp(2i\lambda)h.
\end{equation}
Thus $h$ behaves as a charged scalar field.
Note, however, that the charge of $h$ is two, and therefore the transformation
$\lambda=\pi$ leaves it unchanged. This transformation is the remaining
${\mathbb Z}_2$ symmetry that is left after $h$ gets an expectation value.

Now, we need a gauge invariant expression for the
difference of the Higgs phase angle at neighbouring lattice sites
\begin{eqnarray}
\Delta_i&=&\tr h^\dagger(\x) \tV_{i,-}(\x)h(\xpi)\tV_{i,+}^\dagger(\x)
\nonumber\\
&=&\tr \ch(\x) \Pi_-(\x) U_i(\x) \Pi_-(\xpi) \ch(\xpi)
\Pi_+(\xpi) U_i^\dagger(\x) \Pi_+(\x).
\end{eqnarray}
Defining the Higgs phase angle $\gamma$ by $h\propto\exp(i\gamma(\x))$
and using Eq.~(\ref{equ:gaugedef}), we find
(cf.~Eq.~(10) in Ref.~\cite{Kajantie:1998bg})
\begin{equation}
\delta_i\equiv
\arg\Delta_i=\left[\gamma(\xpi)-\gamma(\x)-2\alpha_i(\x)\right]_\pi
,
\end{equation}
where we use the notation $[X]_\pi\equiv X+2\pi n_X$, with such an $n_X$ that
 $[X]_\pi\in(-\pi,\pi]$.

To define the winding number we have to subtract the magnetic flux. 
This leads to the expression
\begin{equation}
Y_{ij}(x)=\delta_i(x)+\delta_j(\xpi)-\delta_i(\xpj)
-\delta_j(x)-2\alpha_{ij}(\x),
\label{equ:winding}
\end{equation}
which is obviously a multiple of $2\pi$ and thus the winding number
$n_{ij}\equiv(2\pi)^{-1}Y_{ij}$ is always an integer. The value of $Y_{ij}$
is only defined modulo $4\pi$, and therefore one can only distinguish between
even and odd values of $n_{ij}$.
If it is odd,
there is a vortex going through the plaquette.

The definition of the winding number
can be extended to more complicated curves than just single
plaquettes in a straightforward way by adding all the $\delta$'s and
$\alpha_i$'s 
along the curve, and the result is additive
\begin{equation}
n_{C+C'}=n_C+n_{C'}\quad\mbox{mod}\quad 2.
\end{equation}
This equation implies that vortices cannot end.

\subsection{Vortex tension}
Now that we have the expression for the winding number,
we have to restrict the path integral to configurations with
a given winding. The most natural way of doing it is by boundary
conditions.
In lattice simulations, the boundary conditions are almost always chosen
to be periodic,
\begin{equation}
\label{equ:perchi}
\Phi(\vec x+N\jh)=\Phi(\vec x),\qquad
\chi(\vec x+N\jh)=\chi(\vec x),\qquad
U_k(\vec x+N\jh)=U_k(\vec x).
\end{equation}
This implies periodicity also for $\delta_i$ and $\alpha_i$, and therefore
the total winding is even, i.e.~there are
no vortices.

Besides their simplicity, one reason for using the periodic boundary
conditions is that they minimize the finite-size effects, since
they guarantee that the actual boundaries will be invisible to
the physics.
This is particularly important when a defect free energy is
measured, because if there are physical boundaries on the lattice,
their contribution to the free energy can dominate over it.
However, any boundary conditions
that preserve the translation invariance of the system avoid these problems,
and therefore it is enough to have boundary conditions that are periodic
up to any symmetries of the theory.
We can, for instance, modify the boundary conditions (\ref{equ:perchi})
by
\begin{equation}
U_2(x+N,y,z)=\left\{
\matrix{-U_2(x,y,z),\quad\mbox{if}\quad y=1,\cr
U_2(x,y,z),\quad\mbox{if}\quad y\neq 1.}\right.
\label{equ:expbc}
\end{equation}
Changing the sign of $U_2(x,1,z)$ 
does not change the Lagrangian (\ref{equ:lagr2h})
and therefore there are no boundary effects.
Moreover, $\delta_i$ are still periodic, since they always contain $U_i$
as well as $U_i^\dagger$. However, $\alpha_y(x,1,z)$ changes by $\pi$,
and if we calculate the winding of the
whole $xy$ plane according to Eq.~(\ref{equ:winding}), $\alpha_y(x,1,z)$
appears once, and thus the total winding changes from even to odd.
This shows that these non-periodic boundary conditions lead to configurations
with one vortex in $z$-direction.
The free energies in Eq.~(\ref{equ:fediff}) are
thus given by
\begin{equation}
F_n=-\ln Z=-\ln \int {\cal D}U_i{\cal D}\Phi{\cal D}\chi
\exp\left(-\sum_{\x}{\cal L}_{\rm 2H}\right),
\end{equation}
and the subscript $n\in\{0,1\}$ indicates which of the boundary conditions
(\ref{equ:perchi}), (\ref{equ:expbc}) are used.

Next, we have to write $T=\Delta F/L$ in terms
of expectation values. Let us first notice that when we calculate
the action $S=\sum {\cal L}_{\rm 2H}$, the boundary condition 
(\ref{equ:expbc}) only affects the Wilson term. Instead of saying
that we are using non-periodic boundary conditions, we could say that
the boundary conditions are periodic, but the action is changed by
\begin{equation}
\Delta S=\frac{4}{ag^2}
\sum_{z}
\tr U_1(N,1,z)U_2(1,1,z)U_1^\dagger(N,2,z)U_2^\dagger(N,1,z).
\end{equation}
This corresponds to a change in the integration variable in the
path integral, which should be taken into account when measuring
any observable, but since we are now only interested in the
tension, we do not have to worry about that.
The translation invariance of the boundary conditions
implies that we could equally well choose any coordinates $x_0$, $y_0$
and write
\begin{equation}
\label{equ:diracstring}
\Delta S= \frac{4}{ag^2}\sum_{z}\tr U_{12}(x_0,y_0,z).
\end{equation}
This allows us to write
\begin{equation}
\label{equ:tensio}
T=-\lim_{L\rightarrow\infty}\frac{\ln\langle\exp(-\Delta S)\rangle}{L}.
\end{equation}
This form is not, however, 
particularly suitable for numerical simulations,
because the observable $\exp(-\Delta S)$ and the integration weight
$\exp(-S)$ have only very little overlap.
One solution would be to measure derivatives of $T$. For instance,
\begin{equation}
\frac{\partial T}{\partial m_\chi^2}=
a^2N^2\left(\langle\tr\chi^2\rangle_1-\langle\tr\chi^2\rangle_0\right),
\end{equation}
where the subscript indicates whether the expectation value is calculated
in the one-vortex or the zero-vortex ensemble.
this can then be integrated to yield $T$.
The drawback of this method is that it cannot be used to measure
the absolute value of $T$, only its differences.

Another possibility is to introduce a real number $\epsilon\in[0,1]$
and define non-physical ensembles with the action 
$S+\epsilon\Delta S$. We can define
\begin{equation}
F_\epsilon=-\ln\int{\cal D}U_i{\cal D}\Phi{\cal D}\chi
\exp(-S-\epsilon\Delta S),
\end{equation}
and it is then easy to write the tension as
\begin{eqnarray}
T&=&\frac{1}{L}\int_0^1d\epsilon\frac{dF_\epsilon}{d\epsilon}
=\frac{1}{L}
\int_0^1d\epsilon\langle\Delta S\rangle_\epsilon,
\label{equ:integral}
\end{eqnarray}
where $\langle\ldots\rangle_\epsilon$ means an expectation value calculated
using the action $S+\epsilon\Delta S$.
The tension can now be calculated in numerical simulations by 
measuring the integrand of Eq.~(\ref{equ:integral}) at a large number
of different values of $\epsilon$. 
Note that although ensembles with non-integer $\epsilon$ are
not physical, the final result $T$ is.
This method can be improved by using multi-histogram 
techniques~\cite{Kovacs:2000sy,Ferrenberg:1989ui,Hoelbling:2000su}.
An integration scheme in which $\epsilon=1$ but plaquettes are
added to Eq.~(\ref{equ:diracstring}) one by one has also been
suggested recently~\cite{deForcrand:2000fi}.

The shift (\ref{equ:diracstring}) of the action is a 
't~Hooft loop~\cite{'tHooft:1978hy,Mack:1979rq}
 that
is made stable by closing it via the periodic boundary conditions.
Its relation to the twisted boundary conditions (\ref{equ:expbc})
was first discussed in Refs.~\cite{'tHooft:1979uj,Mack:1980kr}.

In pure-gauge SU(2), the vortex tension has been measured 
in Refs.~\cite{Kovacs:2000sy,Hart:2000en,deForcrand:2000fi}.
Instead of having a closed loop, the sum over $z$
in Eq.~(\ref{equ:diracstring}) 
can also be restricted to a shorter interval $\{z_i,\ldots,z_f\}$
\begin{equation}
\langle \bar\mu(z_f)\mu(z_i) \rangle=
Z^{-1}\int{\cal D}U_i{\cal D}\Phi{\cal D}\chi
\exp\left(-S-\frac{4}{ag^2}\sum_{z=z_i}^{z_f}\tr U_{12}(x_0,y_0,z)
\right).
\end{equation}
This technique was used in Ref.~\cite{Hoelbling:2000su}
to study the monopole-antimonopole interaction in the 
pure-gauge SU(2) theory. If the space is interpreted as 2+1 dimensional
and the $z$ direction as time, this describes a creation of a vortex
at time $z_i$ and its annihilation at time 
$z_f$~\cite{Korthals-Altes:2000gs}.
This leads to a dual picture of the theory in which the fundamental
degrees of freedom are vortices
and the phase transition has an interpretation of a spontaneous symmetry
breakdown of a magnetic symmetry for which $\langle\mu\rangle$
defined by
\begin{equation}
\langle\mu\rangle^2=\lim_{z\rightarrow\infty}
\langle \bar\mu(z)\mu(0) \rangle
\end{equation}
is an order parameter.

\section{3D Georgi-Glashow model}
\label{sect:1Higgs}
Let us now consider a system with only one adjoint Higgs field.
We will use the Lagrangian
\begin{eqnarray}
{\cal L}_{\rm 1H}&=&
\frac{4}{ag^2}\sum_{i<j}\left(1-\frac{1}{2}
{\rm Re}\,\tr U_{ij}(\x) 
\right)\nonumber\\
&&+\sum_{i}2a \left[\tr\Phi^2(\x)-\tr\Phi(\x)U_i(\x)
\Phi(\x+i)U_i^\dagger(\x)\right]\nonumber\\
&&+m^2a^3\tr\Phi^2(\x)+a^3\lambda
(\tr\Phi^2)^2.
\label{equ:lagr}
\end{eqnarray}

The topological defects in this system are the 't~Hooft-Polyakov 
monopoles~\cite{'tHooft:1974qc,Polyakov:1974ek},
which are point-like objects. Thus we can define the mass of 
a monopole using Eq.~(\ref{equ:fediff}) simply as
\begin{equation}
\label{equ:massdef}
M=\Delta F.
\end{equation}
Mean field theory predicts that monopoles are massive
in the broken phase and massless in the symmetric phase. Therefore
the mass would be a useful order parameter for the phase transition.
However, the masslessness of monopoles
in the three-dimensional theory
would not imply that there are infinite correlation lengths in the symmetric
phase, because a monopole is a point-like object and therefore
its mass is not related to the correlation length of any operator 
of the dual theory.
In four dimensions, creation and annihilation operators of
monopoles have been
discussed
in Ref.~\cite{Frohlich:1999wq} using a different approach.

\subsection{Boundary conditions}
Again, our strategy is to find boundary conditions that preserve the
translation invariance of the system but force the total
magnetic charge of the lattice to be one.
Periodic boundary conditions 
\begin{equation}
\label{equ:perbc}
\Phi(\vec x+N\jh)=\Phi(\vec x),\qquad
U_k(\vec x+N\jh)=U_k(\vec x),
\end{equation}
are obviously ruled out, because they don't allow non-zero total charge.
Instead, we need more complicated boundary conditions, and we can use
as a guidance
the classical continuum monopole solution
\begin{equation}
\label{equ:monosol}
\Phi(\vec{x})\approx {x_k\sigma_k\over r},\qquad 
A_i(\vec{x})\approx{\epsilon_{ijk}x_j\sigma_k\over 2r^2}.
\end{equation}
If we move from one boundary to another, we
reverse the sign of the coordinate $x_j$, and the fields 
transform as
\begin{equation}
x_j\rightarrow -x_j:\qquad
\Phi\rightarrow -\sigma_j\Phi\sigma_j,\quad
A_i\rightarrow \sigma_j A_i\sigma_j.
\end{equation}
This suggests the boundary conditions
\begin{equation}
\Phi(\vec x+N\jh)=-\sigma_j\Phi(\vec x)\sigma_j,\qquad
U_k(\vec x+N\jh)=\sigma_jU_k(\vec x)\sigma_j.
\label{equ:bc}
\end{equation}
It is straightforward to see that this is a symmetry of the
Lagrangian (\ref{equ:lagr}), and therefore does not break translation
invariance.

\FIGURE{
\epsfig{file=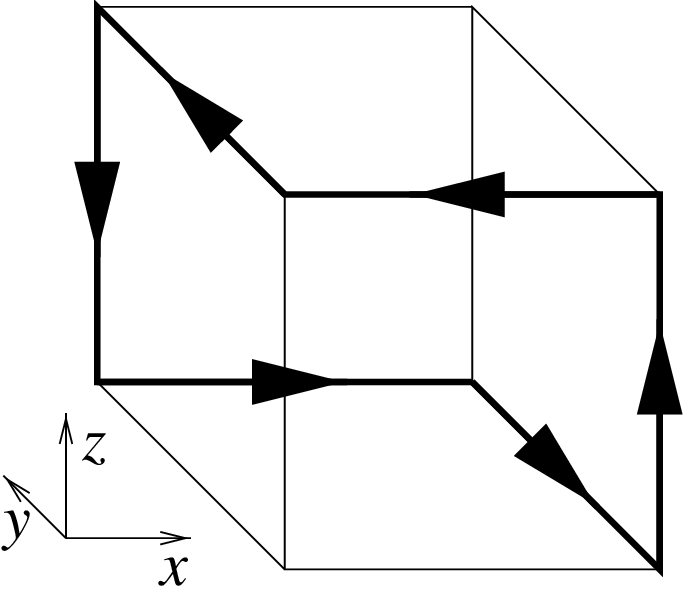,angle=270,width=5.9cm}
\caption{The curve used in Eq.~(\ref{equ:flux})
to
separate the boundary of the lattice into two halves.
}
\label{fig:curve}
}

The boundary conditions (\ref{equ:bc}) imply that
$\Pi_+(\vec x+N\jh)=\sigma_j\Pi_-(\vec x)\sigma_j$
and, consequently,
\begin{equation}
\alpha_{ij}(\vec x+N\hat{k})
=-\alpha_{ij}(\vec x),
\end{equation}
i.e.~the direction of the magnetic flux is reversed at the boundaries.
Thus, when we cross the boundary of the lattice,
we enter a charge-conjugated copy of the same lattice from the
opposite boundary. This is however not enough to guarantee that the total
charge of each of these copies is non-zero.

To determine the total charge, we calculate the magnetic flux through
half of the boundary. The curve defined in Fig.~\ref{fig:curve}
separates the boundary into two halves, and we can calculate
\begin{equation}
\label{equ:flux}
{\rm flux}=\arg\tr\prod_{\rm curve}\tV_{i,+}(\x)
=\arg\prod_{ijk~\rm cyclic}\prod_{n=0}^{N-1}
\tr\tV_{i,+}(n\ih+N\jh)\tV_{i,+}^\dagger(n\ih+N\kh),
\end{equation}
where the final form follows from the fact that
the $\tV_{i,+}$ commute.
To calculate this, we need to know the boundary conditions induced by
Eq.~(\ref{equ:bc}) for $R$:
\begin{eqnarray}
\label{equ:Rbc}
R(\x+N\xh)&=&-\sigma_1R(\x)\sigma_1,\nonumber\\
R(\x+N\yh)&=&-\sigma_2R(\x)\sigma_2,\nonumber\\
R(\x+N\zh)&\propto&\sigma_3R(\x)[\ph(\x),\sigma_3]\sigma_3.
\end{eqnarray}
Then it is easy to calculate the corresponding boundary conditions 
for $\tV_{i,\pm}$:
\begin{eqnarray}
\tV_{i,+}(\x+N\xh)&=&\sigma_1\tV_{i,-}(\x)\sigma_1,\nonumber\\
\tV_{i,+}(\x+N\yh)&=&\sigma_2\tV_{i,-}(\x)\sigma_2,\nonumber\\
\tV_{i,+}(\xh+N\zh)
&\propto&\sigma_3[\sigma_3,\ph(\xh)]\tV_{i,-}(\xh)
[\ph(\xpi),\sigma_3]\sigma_3.
\end{eqnarray}
The case $i=3$ does not contribute, because
\begin{eqnarray}
\label{equ:zcancel}
\tr\tV_{3,+}(n\zh\!+\!N\xh)\tV_{3,+}^\dagger(n\zh\!+\!N\yh)
&=&\tr\sigma_1\tV_{3,-}(n\zh)\sigma_1\sigma_2\tV_{3,-}^\dagger(n\zh)\sigma_2
\nonumber\\
&=&\tr\tV_{3,-}(n\zh)\tV_{3,-}^\dagger(n\zh)=1,
\end{eqnarray}
where we have used the fact that $\sigma_1\sigma_2P_\pm=i\sigma_3P_\pm
=\pm iP_\pm.$
For the horizontal segments we find
\begin{eqnarray}
&&\tr\tV_{1,+}(\x\!+\!N\yh)\tV_{1,+}^\dagger(\x\!+\!N\zh)
\nonumber\\&&\hspace*{2cm}
\propto
\tr \tV_{1,-}(\x)\sigma_1
[\sigma_3,\ph(\x\!+\!\xh)]\tV_{1,-}^\dagger(\x)[\ph(\x),\sigma_3]\sigma_1,
\nonumber\\
&&\tr\tV_{2,+}(\x\!+\!N\zh)\tV_{2,+}^\dagger(\x\!+\!N\xh)
\nonumber\\&&\hspace*{2cm}
\propto
\tr \sigma_2[\sigma_3,\ph(\x)]\tV_{2,-}(\x)[\ph(\x\!+\!\yh),\sigma_3]\sigma_2
\tV_{2,-}^\dagger(\x),
\end{eqnarray}
where we have written $n\ih=\x$ for notational simplicity.
Because
\begin{equation}
\sigma_1[\sigma_3,\ph]=2\left(\matrix{-\ph_1-i\ph_2&0\cr 0&\ph_1-i\ph_2
}\right),
\quad
\sigma_2[\sigma_3,\ph]=2i\left(\matrix{\ph_1+i\ph_2&0\cr 0&\ph_1-i\ph_2
}\right).
\end{equation}
and because $\tV_{i,-}$ projects to the lower right component, we obtain
\begin{eqnarray}
\tr\tV_{1,+}(n\xh+N\yh)\tV_{1,+}^\dagger(n\xh+N\zh)
&\propto&
\exp\left\{-i\left[\theta((n+1)\xh)-\theta(n\xh)\right]\right\},\nonumber\\
\tr\tV_{2,+}(n\yh+N\zh)\tV_{2,+}^\dagger(n\yh+N\xh)
&\propto&
\exp\left\{i\left[\theta((n+1)\yh)-\theta(n\yh)\right]\right\},
\end{eqnarray}
where $\ph_1+i\ph_2\propto\exp(i\theta)$.
In Eq.~(\ref{equ:flux}) all other phases cancel except those at
the corners, and we obtain
\begin{eqnarray}
{\rm flux}&=&
\arg\exp\left\{
i\left[\theta(N\yh)-\theta(N\xh)\right]\right\}=\arg(-1)=\pi,
\end{eqnarray}
where we have used the boundary conditions for $\ph$ in Eq.~(\ref{equ:bc}).

Thus we have shown that with these boundary conditions, there will
be a flux $\pi$ through each of the halves of the boundary, and since
the boundary conditions force the direction of the flux to be
opposite, this means a total flux of $2\pi$ from the lattice. Note that
because the flux is only defined modulo $2\pi$, it is only possible
to force the total magnetic charge
to be either even or odd.

\subsection{Monopole mass}
Using the boundary conditions (\ref{equ:bc}), we can now define the effective
mass of an isolated monopole. 
For any boundary conditions, the free
energy is defined by
\begin{equation}
\label{equ:freeen}
F=-\ln \int DU_iD\Phi \exp\left(-\sum_{\x}
{\cal L}_{\rm 1H}(\x)\right).
\end{equation}
If we denote by $F_0$ and $F_1$ the free
energies for ensembles with boundary conditions (\ref{equ:perbc}) and
(\ref{equ:bc}), respectively, we can use Eq.~(\ref{equ:massdef})
directly to define the monopole mass $M$.

In order to write $M$ as an expectation value,
we will now transform the
one-monopole system in such a way that the difference is moved from
the boundary conditions into a shift in the action.
Let us
consider a field configuration that satisfies the
conditions (\ref{equ:bc}), and apply the following (large)
gauge transformation
\begin{equation}
\label{equ:gaugetrans}
\Lambda=\left\{\begin{array}{ll}
1,& {\rm if}~x,z<N\\
i\sigma_1,&{\rm if}~x<N,~z\ge N\\
i\sigma_3,&{\rm if}~x\ge N,~z<N\\
i\sigma_2,&{\rm if}~x,z\ge N
\end{array}\right..
\end{equation}
This will change the boundary conditions into
\begin{equation}
\label{equ:ccbc}
\Phi(\x+N\jh)=-\sigma_2\Phi(\x)\sigma_2=\Phi^*(\x),\qquad
U_k(\x+N\jh)=\sigma_2U_k(\x)\sigma_2=U_k^*(\x),
\end{equation}
everywhere else except at the edges of the lattice, where
\begin{eqnarray}
U_3(x,N,N-1) &=& -U_3{}^*(x,0,N-1)
,\nonumber\\
U_1(N-1,N,z) &=& -U_1{}^*(N-1,0,z)
,\nonumber\\
U_1(N-1,y,N) &=& -U_1{}^*(N-1,y,0)
.
\end{eqnarray}
The C-periodic boundary conditions (\ref{equ:ccbc}) have been
discussed before in Ref.~\cite{Kronfeld:1991qu} in a
different context.

When calculating the free energy (\ref{equ:freeen}),
we can change the boundary conditions everywhere to Eq.~(\ref{equ:ccbc})
by redefining
\begin{eqnarray}
\label{equ:redef}
U_3(x,N,N-1) &\rightarrow&-U_3(x,N,N-1)
,\nonumber\\
U_1(N-1,N,z) &\rightarrow&-U_1(N-1,N,z)
,\nonumber\\
U_1(N-1,y,N) &\rightarrow&-U_1(N-1,y,N)
.
\end{eqnarray}
This changes the Wilson term in 
in the Lagrangian (\ref{equ:lagr}). 
The total effect 
is to flip the sign of the Wilson term at three of the edges,
and consequently the one-monopole free energy is given by
\begin{equation}
\label{equ:freeen1}
F_1=-\ln \int DU_i D\Phi \exp\left(-\sum_{\x}
{\cal L}_{\rm 1H}(\x)-\Delta S\right),
\end{equation}
with C-periodic
boundary conditions~(\ref{equ:ccbc}). 
The change of the action is
\begin{eqnarray}
\Delta S&=&
\frac{4}{ag^2}{\rm Re}\left(
\sum_{x=0}^{N-1}
\tr 
U_{23}(x,N-1,N-1)
\right.\nonumber\\
&&\left.
+
\sum_{y=0}^{N-1}
\tr 
U_{13}(N-1,y,N-1)
+
\sum_{z=0}^{N-1}
\tr 
U_{12}(N-1,N-1,z)
\right).
\end{eqnarray}

Because these boundary conditions
 preserve translation
invariance, we could as well choose any $\x_0=(x_0,y_0,z_0)$ and
write
\begin{eqnarray}
\label{equ:monodeltaS}
\Delta S&=&
\frac{4}{ag^2}{\rm Re}\left(
\sum_{x=0}^{N-1}\tr U_{23}(x,y_0,z_0)
\nonumber\right.\\
&&\left.
+
\sum_{y=0}^{N-1}\tr U_{13}(x_0,y,z_0)
+
\sum_{z=0}^{N-1}\tr U_{12}(x_0,y_0,z)\right).
\end{eqnarray}
Note how similar this expression is to Eq.~(\ref{equ:diracstring}).
The only difference is that we have three intersecting 't~Hooft loops 
instead of just one.
Again, we 
emphasize that, because Eq.~(\ref{equ:freeen1}) is equivalent to
Eq.~(\ref{equ:freeen}) with the 
translation invariant boundary conditions Eq.~(\ref{equ:bc}),
the choice of $\x_0$ does not affect any 
observable,
and in particular, it does not fix the location of the monopole
on the lattice.

Without $\Delta S$, Eq.~(\ref{equ:freeen1}) would be precisely
the ordinary free energy (\ref{equ:freeen}) with 
C-periodic boundary
conditions~(\ref{equ:ccbc}). 
As they obviously imply
\begin{equation}
\tV_{i,+}(\x+N\jh)=\sigma_2\tV_{i,-}(\x)\sigma_2,
\end{equation}
the flux through the curve in Fig.~\ref{fig:curve}
vanishes~(cf.~Eq.~(\ref{equ:zcancel})). Consequently, 
we can instead of the periodic boundary conditions (\ref{equ:bc}),
use the boundary conditions in Eq.~(\ref{equ:ccbc}) to define $F_0$.
This leads to the expression
\begin{equation}
\label{equ:3dmassdef}
M=-\ln\langle\exp(-\Delta S)\rangle,
\end{equation}
where $\Delta S$ is given by Eq.~(\ref{equ:monodeltaS})
and the boundary conditions by Eq.~(\ref{equ:ccbc}).

To measure $M$,
we can use the  
derivative with respect to $m^2$,
\begin{equation}
\label{equ:Mderiv}
\frac{\partial M}{\partial m^2}=V\left(
\langle \tr\Phi^2\rangle_1-
\langle \tr\Phi^2\rangle_0\right),
\end{equation}
where the subscript tells whether the expectation value is calculated in
the zero-monopole or one-monopole system, and $V=a^3N^3$ is the volume of
the system.
If we assume that at large enough $m^2$, i.e.~in the symmetric phase,
the monopole mass vanishes, we only have to measure $\partial M/\partial m^2$
at sufficiently many values of $m^2$ and integrate.

Like in the case of a vortex, we can also define for $\epsilon\in[0,1]$
\begin{equation}
\label{equ:freeeneps}
F_\epsilon=-\ln \int DU_i D\Phi\exp\left(-\sum_{\x}
{\cal L}_{\rm 1H}(\x)-\epsilon\Delta S\right).
\end{equation}
Using Eq.~(\ref{equ:freeeneps}), we can now write the monopole mass as
\begin{equation}
\label{equ:monomass}
M=\int_0^1d\epsilon\frac{\partial F_\epsilon}{\partial\epsilon}
=\int_0^1d\epsilon\langle\Delta S\rangle_\epsilon,
\end{equation}
where the subscript $\epsilon$ indicates that the expectation value must
be measured in the presence of the insertion $-\epsilon\Delta S$ as in
Eq.~(\ref{equ:freeeneps}).
This gives us the absolute value of $M$, but with the cost that we have
to measure expectation values at non-physical values of $\epsilon$.

\section{4D pure gauge SU(2)}
\label{sect:puregauge}
Because of the large amount of literature on condensation of monopoles
as an explanation of confinement in pure-gauge SU($N$) 
theory~\cite{DelDebbio:1995sx,Chernodub:1997ps,Mandelstam:1976pi,%
Cea:2000zr,Kronfeld:1987vd}, it is
interesting to see whether our formulation can be applied to this case
as well.

The Lagrangian of the 4D pure gauge theory is
\begin{equation}
{\cal L}_{\rm 4D}=
\beta\sum_{\mu<\nu}\left(1-\frac{1}{2}
{\rm Re}\,\tr U_{\mu\nu}(\x) 
\right).
\label{equ:lagr4d}
\end{equation}
In order to define monopoles, we will also have to specify the 
Abelian projection~\cite{'tHooft:1981ht} we are using, 
i.e.~how $\Phi$ is related to the gauge fields. 
The original suggestion by 't~Hooft~\cite{'tHooft:1981ht}
was to
choose an operator $\O$ that is a product of the link variables
and to define $\O=\O_0+i\Phi$,
where $\O_0$ is proportional to the unit matrix.

In order to use our definition for the monopole mass, the boundary 
conditions must be chosen in such a way that the 
induced boundary condition for $\Phi$ is compatible with
Eq.~(\ref{equ:bc}). 
However, for most choices of $\O$ this is not
possible at all.
This is obviously the case with all definitions of $\O$ that are
products of spatial links $U_i$, because then the
boundary conditions~(\ref{equ:bc}) imply
\begin{equation}
\O(t,\x+N\ih)=\sigma_i\O(t,\x)\sigma_i,\quad
{\rm i.e.}\quad
\Phi(t,\x+N\ih)=\sigma_i\Phi(t,\x)\sigma_i,
\end{equation}
in contradiction with Eq.~(\ref{equ:bc}). 
As a consequence, these Abelian projections 
are also
incompatible with the classical monopole solution (\ref{equ:monosol}).

There are suitable ways of defining $\Phi$, though,
and one of them is the most popular choice of an Abelian projection,
namely the maximally Abelian gauge~\cite{'tHooft:1981ht,Kronfeld:1987vd}.
It means finding the gauge transformation $\Lambda(t,\x)$ 
that maximizes the functional
\begin{equation}
\label{equ:maxab}
R_{\rm MAG}=
\sum_{\mu,t,\x}\tr \sigma^3\tU_\mu(t,\x)\sigma^3\tU_\mu^\dagger(t,\x),
\end{equation}
where $\tU_i(t,\x)=\Lambda^\dagger(t,\x) U_i(t,\x) \Lambda(t,\xpi)$
and $\tU_0(t,\x)=\Lambda^\dagger(t,\x) U_0(t,\x) \Lambda(t+1,\x)$.
Defining $\ph(t,\x)=\Lambda(t,\x)\sigma^3\Lambda^\dagger(t,\x)$,
we can write Eq.~(\ref{equ:maxab}) as
\begin{eqnarray}
R_{\rm MAG}&=&
\sum_{t,\x}\Biggl[
\tr\ph(t,\x)U_0(t,\x)\ph(t\!+\!1,\x)U_0^\dagger(t,\x)
\nonumber\\&&\phantom{\sum_{t,\x}\Biggl[}
+\sum_i\tr\ph(t,\x)U_i(t,\x)\ph(t,\xpi)U_i^\dagger(t,\x)
\Biggr],
\end{eqnarray}
which must be maximized with respect to the field $\ph(t,\x)$.
It is obvious that if $\ph$ maximizes $R_{\rm MAG}$
in the region $\x\in\{0,\ldots,N-1\}^3$ and $U_\mu$ and $\ph$ satisfy
the generalization of the boundary conditions (\ref{equ:bc}),
\begin{equation}
\ph(t,\vec x+N\jh)=-\sigma_j\ph(t,\vec x)\sigma_j,\qquad
U_\mu(t,\vec x+N\jh)=\sigma_jU_\mu(t,\vec x)\sigma_j,
\label{equ:mabc}
\end{equation}
then $\ph$ maximizes $R_{\rm MAG}$ everywhere.
Thus the maximally Abelian gauge is compatible
with the boundary conditions~(\ref{equ:bc}).

We will now assume that $\Phi$ is given either by the maximally Abelian
gauge condition or some other way that is compatible with
Eq.~(\ref{equ:mabc}).
The definition of the magnetic field can be 
generalized from Eq.~(\ref{equ:fluxdef}) trivially by
simply using the same definition at every time slice.
Consequently, we can 
perform the gauge transformation (\ref{equ:gaugetrans}) and the
field redefinition (\ref{equ:redef}) or equivalently
simply take Eq.~(\ref{equ:3dmassdef}) and write
\begin{equation}
\label{equ:4dmassdef}
M=-\lim_{L_t\rightarrow\infty}\frac{\ln\langle\exp(-\Delta S)\rangle}{L_t},
\end{equation}
where the boundary conditions in the spatial directions
are C-periodic (\ref{equ:ccbc}),
$L_t$ is the length of the lattice in the time direction
and
\begin{eqnarray}
\Delta S(t)&=&
\beta{\rm Re}\left(
\sum_{x=0}^{N-1}\tr U_{23}(t,x,y_0,z_0)
\right.\nonumber\\&&\left.
+
\sum_{y=0}^{N-1}\tr U_{13}(t,x_0,y,z_0)
+
\sum_{z=0}^{N-1}\tr U_{12}(t,x_0,y_0,z)
\right).
\label{equ:deltaS}
\end{eqnarray}
As before, the mass can be measured in practice using the multi-histogram 
techniques~\cite{Kovacs:2000sy,Ferrenberg:1989ui,Hoelbling:2000su}
or plaquette by plaquette~\cite{deForcrand:2000fi}.
Note that Eq.~(\ref{equ:4dmassdef}) 
does not refer to $\Phi$ at all and therefore, 
the tedious task of numerically maximizing $R_{\rm MAG}$
is avoided altogether.
Moreover, as long as the definition of $\Phi$ is compatible with 
Eq.~(\ref{equ:mabc}), the mass of a monopole is independent
of the precise choice of the Abelian projection.

Let us now compare our prescription with that used in
Ref.~\cite{DelDebbio:1995sx}.
There, instead of the monopole mass, the disorder parameter 
$\langle\mu\rangle$ was measured from the correlator
$\langle\bar{\mu}(t_i)\mu(t_f)\rangle$.
The correlator was defined by shifting 
the vertical plaquettes $U_{i0}$ in the action at $t=t_i$ 
\begin{eqnarray}
U_{i0}(t_i,\x)&=&U_i(t_i,\x)U_0(t,\xpi)U_i^\dagger(t_i+1,\x)
U_0^\dagger(t_i,\x)
\nonumber\\
&\rightarrow&
U_i(t_i,\x)U_0(t_i,\xpi)U_i^\dagger(t_i+1,\x)
U_{{\rm cl},i}^\dagger(t_i+1,\x)
U_0^\dagger(t_i,\x),
\end{eqnarray}
where $U_{{\rm cl},i}$ is the gauge field of a static classical Abelian
monopole, and correspondingly at $t=t_f$.
By redefining the spatial links $U_i$,
\begin{equation}
\label{equ:clredef}
U_{{\rm cl},i}(t,\x)U_i(t,\x)\rightarrow U_i(t,\x),
\end{equation} 
the shift can be moved to the
spatial plaquettes, thus making direct comparison possible.
Leaving aside the problems caused by the fact that the redefinition
(\ref{equ:clredef}) changes the boundary conditions
and with most choices of the Abelian projection also the value of $\Phi$,
we can write the resulting shift in the
action as
\begin{equation}
\label{equ:cldeltaS}
\Delta S(t)=-\beta\sum_{\x}\sum_{i<j}{\rm Re}\left(\frac{1}{2}\tr 
U'_{ij}(t,\x)-\frac{1}{2}\tr U_{ij}(t,\x)
\right),
\end{equation}
where $U'_i=U_{{\rm cl},i}^\dagger U_i$.

The structure of Eqs.~(\ref{equ:deltaS}) and (\ref{equ:cldeltaS})
is similar, and realizing that in the full non-perturbative treatment there
is no reason for $U_{{\rm cl},i}$ to be a solution of the classical
field equations, we can interpret Eq.~(\ref{equ:deltaS})
as simply a particularly convenient choice of $U_{{\rm cl},i}$.
For instance, it makes it much easier to choose suitable boundary conditions
that preserve the translation invariance and allow a non-zero total
magnetic charge.

The measurement of the monopole mass has also been discussed in
Ref.~\cite{Cea:2000zr}. There the non-zero
total magnetic charge was achieved by using fixed boundary conditions.
However, this breaks the translation invariance,
and boundary effects can therefore be significant. 

\section{Conclusions}
In this paper,
we have presented a way of measuring 
free energies of topological defects in SU(2) gauge theories on
a lattice.
Our approach was to impose ``twisted'' boundary conditions,
which preserve the translation invariance and minimize boundary
effects but guarantee that the number of topological defects inside
the system is non-zero. By changing the integration variables in the
partition function, the free energy of the defect can then be
expressed as an expectation value, which makes it possible to measure
it in numerical simulations.
In the case of the two-Higgs model, we derived 
an expression for the vortex tension in
terms of a 't~Hooft loop using this technique.

The main result of this paper is the derivation of an analogous
expression for
the mass of a 't~Hooft-Polyakov monopole in the Georgi-Glashow model,
defined as the free energy difference between configurations of
total magnetic charge one and zero. 
Again, the construction
was based on manifestly translation invariant
boundary conditions, which guarantees that the boundary effects are 
minimal. By a gauge
transformation and a redefinition of the fields, we rewrote
the free energy as a path integral with 
C-periodic
boundary conditions and with an insertion $\Delta S$, which 
consists of three intersecting 't~Hooft loops.
In addition to avoiding problems with boundary conditions, our
prescription leads to a simpler final expression for the
monopole mass than the ones appeared previously in the literature.

We believe that the monopole mass (\ref{equ:3dmassdef})
can be used to distinguish between 
the
phases of the Georgi-Glashow model, because it is zero in the SU(2) 
phase and non-zero
in the U(1) phase. 
More generally, we conjecture that if
the symmetry breaking structure of a phase transition in a gauge
theory allows topological defects and it is
possible to construct an observable that measures the free energy of
such a defect in the same way as Eqs.~(\ref{equ:tensio}) and 
(\ref{equ:3dmassdef}) measure the free
energies of vortices and monopoles, the point at which the defect
free energy becomes non-zero defines a transition point between the
phases. Although this does not necessarily imply a singularity
in the partition function, it suggests that in these cases the
phase transition is not likely to be a smooth crossover.

We also showed that the same construction can be used in 
pure-gauge theory to
observe condensation of Abelian monopoles.
In this framework, it is not necessary to fix the gauge or carry out
the Abelian projection in the simulation. This simplifies
the procedure significantly and means that the observable, and 
therefore
also the resulting monopole mass, is manifestly independent of the
choice of the Abelian projection, provided
that it is compatible with 
the classical 't~Hooft-Polyakov monopole solution.
Thus, the ambiguities related to the differences of 
various Abelian projections can be avoided
by adopting Eq.~(\ref{equ:4dmassdef}) as the definition of condensation of
Abelian monopoles.

\acknowledgments{
The authors wish to thank K. Rummukainen for generously 
providing the code which acted as a template
for the two Higgs calculations. This work was supported in part by PPARC
and an ESF network.
H.P.S.~was supported by the JSPS and A.R.~partially by
the University of Helsinki. The computational work for
this 
paper was carried out on a Silicon Graphics Origin 2000 supercomputer of
the 
UK Computational Cosmology Consortium based at the University of
Cambridge.}

\appendix
\section{Computational Details of the 
Two Higgs model calculation}
The configurations for this calculation were generated using a combination
of the Kennedy-Pendleton \cite{Kennedy:1985nu} heatbath algorithm and the 
over-relaxed 
\cite{Creutz:1987xi,Brown:1987rr}  algorithm 
along with a Metropolis accept-reject step for the pure gauge sector and 
the scalar-gauge interactions and a Metropolis step for the scalar interactions. 
Measurements are carried out every 500 compound sweeps. In 
Table~\ref{table:runs}, we list the 
thermalization and number of measurements carried out for the lattices 
used in Fig.~\ref{fig:hugh}.

\TABLE[ht]{
\begin{tabular}{|c|c|c|}\hline
$V/a^3$ & Number of thermalization steps & Number of measurements \\ \hline
$12^3$ & 50,000 & 200 \\ \hline
$16^3$ & 50,000 & 200 \\ \hline
$24^3$ & 10,000 & 40 \\ \hline
\end{tabular}
\caption{Run parameters for lattice sizes presented in Fig.~1.}
\label{table:runs}
}

As outlined in \cite{Hart:1997ac},
if we define 
\begin{equation}
\rho_{(i)}(\x) =   a \tr \chi_{(i)}^2 ,
\end{equation}
where $i$ indicates the $i$-th configuration, (the fields computed on the lattice 
are $\sqrt{a} \chi$) and 
\begin{equation}
\overline{\rho_i} = \frac{V}{a^3} \sum_{\x} \rho_{(i)}(\x) ,
\end{equation}
then the susceptibility in Fig.~\ref{fig:hugh} is defined
as 
\begin{equation}
S(\chi) = \frac{V}{a^3} \left( 
\frac{\langle \overline{\rho_i} \overline{\rho_i} \rangle}
{\langle  \overline{\rho_i}\rangle^2} - 1 \right) ,
\end{equation}
where the notation $\langle {\cal {O}} \rangle$ represents an average over 
configurations.

\end{document}